\documentclass[twocolumn,aps,prl,superscriptaddress,floatfix]{revtex4-2}

\usepackage{amsmath}
\usepackage{graphicx}
\usepackage{physics}
\usepackage{bbold}
\usepackage{color}
\usepackage{cancel}

\usepackage[colorlinks=true, allcolors=blue]{hyperref}

\begin{document}

\title{Structural symmetry effects on the competition of density waves and superconductivity in bilayer nickelates}

\author{Steffen B\"otzel}
\altaffiliation{These authors contributed equally to this work.}
\affiliation{Theoretische Physik III, Fakult\"at für Physik und Astronomie, Ruhr-Universit\"at Bochum,
  D-44780 Bochum, Germany}

\author{Aiman Al-Eryani}
\altaffiliation{These authors contributed equally to this work.}
\affiliation{Theoretische Physik III, Fakult\"at für Physik und Astronomie, Ruhr-Universit\"at Bochum,
  D-44780 Bochum, Germany}

\author{Jun Zhan}
\affiliation{Beijing National Laboratory for Condensed Matter Physics and Institute of Physics, Chinese Academy of Sciences, Beijing 100190, China}
\affiliation{School of Physical Sciences, University of Chinese Academy of Sciences, Beijing 100190, China}

\author{Xianxin Wu}
\affiliation{Institute for Theoretical Physics, Chinese Academy of Sciences, Beijing, China}

\author{Frank Lechermann}
\affiliation{Theoretische Physik III,  Fakult\"at für Physik und Astronomie, Ruhr-Universit\"at Bochum,
  D-44780 Bochum, Germany}

\author{Michael M. Scherer}
\affiliation{Theoretische Physik III,  Fakult\"at für Physik und Astronomie, Ruhr-Universit\"at Bochum,
  D-44780 Bochum, Germany}

\author{Ilya M. Eremin}
\affiliation{Theoretische Physik III,  Fakult\"at für Physik und Astronomie, Ruhr-Universit\"at Bochum,
  D-44780 Bochum, Germany}

\pacs{}
\begin{abstract}
We investigate the interplay between spin-density-wave (SDW) order and superconductivity in the bilayer nickelate La$_3$Ni$_2$O$_7$ using the functional renormalization group~(fRG) applied to multiorbital weak-coupling models of both the ambient- and high-pressure crystal structures. As Hund's coupling increases, the leading instability evolves from superconductivity to an SDW state with ordering vector $\mathbf{Q}_1 \approx (\pi/2,\pi/2)$ (equivalently $\mathbf{Q}_Y \approx (0,\pi)$ in the orthorhombic $Amam$ unit cell), in agreement with experimental observations. Surprisingly, the ambient- and high-pressure structures exhibit nearly identical non-interacting susceptibilities and leading fRG instabilities, indicating that the emergence of superconductivity under pressure cannot be explained solely by changes in the low-energy electronic structure. Instead, our results identify the suppression of orthorhombicity as a key ingredient for superconductivity. As the system approaches the tetragonal limit, symmetry-related SDW fluctuations become nearly degenerate, frustrating long-range magnetic order while enhancing pairing interactions. These findings highlight lattice symmetry as a central tuning parameter of the competing ordered states in bilayer nickelates and suggest that reducing orthorhombicity through uniaxial strain could stabilize bulk superconductivity already at ambient pressure.
\end{abstract}

\maketitle
\section{Introduction}

The recent discovery of high-temperature superconductivity in bulk La$_3$Ni$_2$O$_7$ under high pressure~\cite{sun23,JunHou:117302,zhang2023high,zhou2023evidence,zhang2023effects,wang2024structure,wang2023pressure,dong2023visualization} and related systems -- ranging from (La,R)$_3$Ni$_2$O$_7$~\cite{WangNature24,li2026bulk,qiu2025interlayer} with $R$ denoting other rare earths (often $R =Pr,Nd,Sm$) to compressively strained (La,R)$_3$Ni$_2$O$_7$ thin films~\cite{ko2024signatures,zhou2025ambient,bhatt2025resolving,liu2025superconductivity} -- makes the bilayer nickelate an interesting platform to better understand high-temperature superconductivity. 

The bilayer nickelates have a remarkably large interlayer coupling, and both Ni $e_g$ orbitals are located near the Fermi level. The resulting complex low-energy structure has multiple Fermi surface sheets and gives rise to competing orders. The details of the superconducting phase are hotly debated. A plethora of theoretical proposals have been made from both weak and strong coupling frameworks, which have been summarized in recent review articles \cite{wang2024normal,zhang2026superconductivity, pan2026superconductivity, oh2026doping,wang2026interlayer}. Currently, the primary candidates for the superconducting gap symmetry are $s_\pm$ ($A_{1g}$) and $d$-wave types ($B_{1g}$ or $B_{2g}$). Existing experiments have yet to yield a consensus, often providing partially conflicting insights. Point-contact spectroscopy on bulk samples~\cite{cao2025direct,liu2025andreev,guo2025revealing} and ARPES on thin films~\cite{Shen-Chen-2025-ARPES-S+-,li2025angle,wang2025electronic,li2026three} offer different interpretations, while recent STM measurements reveal two superconducting gaps associated with different bands and are compatible with a multiband $s$-wave scenario~\cite{Hai-Hu-Wen-2025-STMSpm,liang-2026-STM-LaPr-Thin-UShape,wang-2026-STM-LaPr-Thin-UShape2Peaks}. 

In the ambient-pressure phase, the bulk samples host a spin-density-wave (SDW) but no superconductivity~\cite{chen2024electronic,chen2026nature}. Applying pressure slightly increases the magnetic ordering temperature~\cite{khasanov2025pressure}. However, the SDW signatures fade away when superconductivity sets in~\cite{fan2026evolution}.

Furthermore, pressure typically leads to a structural transition. At ambient pressure, most works report orthorhombic $Amam$ or monoclinic $P2_1/a$ \cite{li2026bulk} space groups and a transition to the tetragonal $I4/mmm$ space group under high pressure with a possible intermediate orthorhombic $Fmmm$ phase~\cite{wang2024structure}. The emergence of superconductivity was found to (nearly) coincide with the transition to $I4/mmm$ phase \cite{sun23,zhang2025identifying,li2026bulk}.  Interestingly, a tetragonal crystal structure has also been achieved in La$_3$Ni$_2$O$_{6.92}$ under ambient pressure, but in those samples both, the SDW and superconductivity were absent up to pressures of 70 GPa \cite{shi2025spin}, which indicates a relation of both phases. 

On the structural side, higher symmetry and straight Ni-O-Ni bond angles are argued to be the best for superconductivity. One argument is that straight Ni-O-Ni bond angles support superexchange processes via the oxygen $p$ orbitals, thereby increasing the effective interlayer spin-spin interaction $J_{\perp}$ (or effective in-plane spin-spin interaction $J_{\parallel}$ for in-plane Ni-O-Ni bonds). However, it is not clear why this supports superconductivity and not just the SDW phase. 

There is a belief that a Ni-3$d_{z^2}$ dominated band ($\gamma$ band) might be lifted from below to crossing the Fermi level~\cite{liu2023s,wang2024normal,sun23}. This Lifshitz transition could then alter the low-energy environment in a way that it supports superconductivity over a SDW phase~\cite{yang2023possible,jiang2025theory}. This idea is supported by recent RPA calculations indicating that superconducting $T_{\text{c}}$ is highest when the $\gamma$ band gently touches the Fermi level~\cite{ryee2025superconductivity,le2025opposite}. If it further moves upwards, strong ferromagnetic fluctuations act against the formation of singlet superconductivity. In contrast, a recent FLEX study suggest that the $s_\pm$ phase is robust and insensitive to the exact position of the $\gamma$ pocket~\cite{ushio2025theoretical}. ARPES measurements do not yield a conclusive picture on the location of this band \cite{nie2026superconductivity,li2025angle,yang2024orbital,au2025universal,wang2025electronic,li2024electronic}. Recent work points out that the bilayer structure manifests itself in a strong $k_z$ (photon energy) dependence of the important matrix elements, which potentially could explain the differences in earlier measurements and suggests the presence of the $\gamma$~band at the Fermi level in thin films~\cite{li2026three}. 

Anyway, irrespective of the position of the $\gamma$ band, the connection between superconductivity and the SDW phase has not yet been explored in detail. 

In this work, we compare the ambient and high-pressure phases and analyze the leading superconducting and SDW instabilities from the  weak-coupling perspective. Treating Hund's coupling and on-site energy differences $\epsilon_{x^2-y^2}-\epsilon_{z^2}$ as parameters, we show that ambient- and high-pressure phases are extremely similar regarding not only their non-interacting susceptibilities but even the leading instabilities using functional renormalization group~(fRG) calculations. We unravel that there is a low $J$ regime, in which only superconductivity is found and strongly depends on the location of the $\gamma$ band, whereas in the high $J$ regime, the location of the $\gamma$ band has less impact on when the transition between superconductivity and a SDW phase takes place. 

We then address the role of orthorhombicity for the transition by hypothetical application of uniaxial strain. Our results suggest that the emergence of superconductivity in the high-pressure phase is significantly driven by the suppression of orthorhombicity. In particular, we argue that reduced orthorhombicity leads to doubling the number of SDW order parameter components relevant at low energies, which both enhance overall magnetic fluctuations available as ``pairing glue'' for superconductivity, while simultaneously preventing the stabilization of a full-fledged SDW phase.

\section{Model}
We model bulk La$_3$Ni$_2$O$_7$ using an effective two-orbital bilayer model. The non-interacting Hamiltonian reads
\begin{equation}
    \begin{split}
        \mathcal{H}_0=\sum_{\bf{k}\ell\ell'\mu\nu\sigma} H_{0,\ell\mu;\ell'\nu}({\bf{k}}) c_{\bf{k}\sigma\ell\mu}^\dagger c_{\bf{k}\sigma\ell'\nu}^{\phantom{\dagger}},
    \end{split}
\end{equation}
where $c_{\bf{k}\sigma\ell\mu}^\dagger$ creates an electron with momentum~${\bf k}$, spin~$\sigma$, in layer~$\ell$ and effective Ni-$3d-e_g$ orbital~$\mu$. The matrix elements of the tight-binding Hamiltonian $H_{0,\ell\mu;\ell'\nu}(\bf{k})$ are from a projection of the low-energy band structure from density functional theory (DFT) to maximally-localized Wannier states \cite{marzari12}. 

By varying the relative on-site potentials $\delta \epsilon \equiv \epsilon_{x^2-y^2}-\epsilon_{z^2}$, with the chemical potential adjusted to keep the filling fixed, the position of the $d_{z^2}$-dominated $\gamma$ pocket can be shifted relative to the Fermi level. Throughout the manuscript we will focus on three distinct scenarios. The first scenario is that the $\gamma$ pocket touches the Fermi level. The corresponding value for $\delta \epsilon$ is denoted by $\delta \epsilon^*$. The other two scenarios are denoted by the reference values $\delta \epsilon^{+/-}$ corresponding to the case where the top of the $\gamma$ pocket is 50 meV above/below the Fermi level.

\begin{figure}
	\includegraphics[width=1\linewidth]{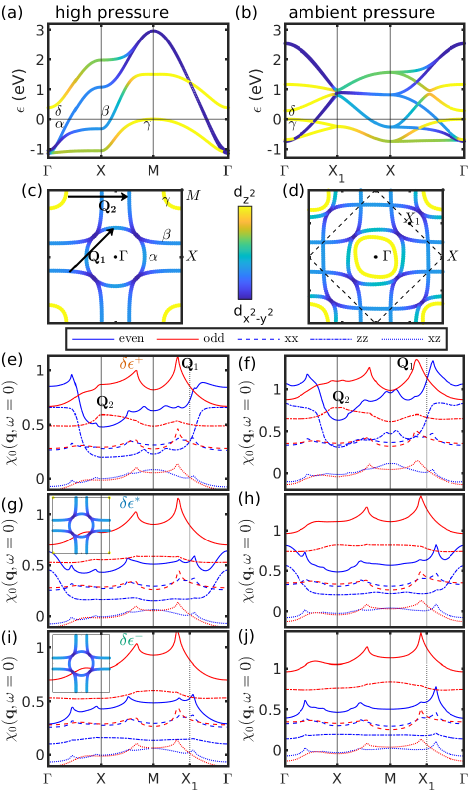}
	\caption{Band structure, Fermi surface, and non-interacting susceptibility of La$_3$Ni$_2$O$_7$ at high and at ambient pressure. (a)~Band structure for the high-pressure model along high symmetry path for $\delta\epsilon = \delta\epsilon^*$. (b)~Band structure for the ambient-pressure model along the $Amam$ high-symmetry path.  Fermi surface for high- and ambient-pressure phases for $\delta\epsilon = \delta\epsilon^+$ ~(c,d). Total and orbital-resolved contributions to the non-interacting susceptibilities for the top of the $\gamma$~band being located 50 meV above (e,f), at (g,h), and 50 meV below (i,j) the Fermi level using the high-symmetry path. (e,g,i)~are for the high-pressure symmetry, whereas (f,h,j) are for the ambient-pressure model. Similarities in band structures and bare susceptibilities are detailed in the text.}
\label{Fig:1}
\end{figure}

We consider three distinct models. The first two represent unaltered \textit{ab initio} calculations of two physical situations: (1)~A model in the high-pressure phase (29.5 GPa) which adopts a $Fmmm$ space group with almost nearly crystal structure, see left column of Fig.~\ref{Fig:1}, and (2)~A model in the ambient-pressure phase realizing an orthorhombic $Amam$ space group, see right column of Fig.~\ref{Fig:1}. (3)~The remaining model explores a theoretical scenario with modeled uniaxial strain at ambient pressure conditions. In particular, we enforce $a = b$ with allowed structural relaxation. For all of these models, we consider the free-standing limit, where inter-bilayer couplings are neglected, which allows an effective two-dimensional description. \texttt{Wannier90} files of these models are made available in the Supplemental Material.

In the $Fmmm$ phase, there is one nickel per layer in the unit cell so that there are in total two nickel atoms per unit cell. In contrast, in the $Amam$ phase, there are two nickel atoms per layer and four in total per unit cell. The doubling of the unit cell for the $Amam$ phase comes with folding the Brillouin zone compared to the $Fmmm$ phase, which doubles the number of bands, cf. Fig.~\ref{Fig:1}. The hopping parameters for the high-pressure case are taken from Ref.~\cite{lechermann2023electronic} and for the ambient pressure model from Ref.~\cite{khasanov2025pressure}.

For the interaction part of the Hamiltonian, on-site intraorbital ($U$), interorbital ($U'$), Hund's type ($J$), and pair hopping ($J'$) interactions are included, assuming spin-rotational invariance by using $U' = U - 2J$ and $J = J'$ \cite{graser2009near}. Non-on-site interactions are neglected.

\section{Method}
To investigate the competition between the SDW and superconducting phase in our models, we employ the fermionic fRG approach, which accounts for pairing, charge, and spin fluctuations and ordering tendencies on equal footing~\cite{10.1143/PTP.105.1,RevModPhys.84.299,Platt_2013,Dupuis:2020fhh}. We focus on the evolution of the two-particle interaction~ $\Gamma^{(4)}$ with respect to a sliding energy scale~$\Lambda$, which interpolates between the microscopic model interaction and the low-energy effective interaction in which fluctuations have been integrated out. An instability is identified by the divergence of $\Gamma^{(4)}$ at a scale $\Lambda_c$, which indicates a tendency of the normal state to an ordered symmetry-broken state. 

In the truncated-unity fRG scheme~\cite{Lichtenstein_2017,Profe_2022_tu2frg,Beyer_2022,Profe_2024_diverge,10.21468/SciPostPhysCodeb.26-r0.5}, the vertex is decomposed into a sum of three scattering channels: crossed particle-hole~$C$, direct particle-hole~$D$ and particle-particle~$P$, each of which depends most strongly on one of the three Mandelstam momentum-frequency variables~$q = (i\Omega, \mathbf{q})$~\cite{PhysRevB.79.195125}. We employ a standard approximation for the instability analysis, neglecting frequency dependencies, which are RG~irrelevant~\cite{10.1143/PTP.105.1,RevModPhys.84.299}. Secondary momentum and orbital dependencies beyond the Mandelstam variables are expanded in a basis of form-factors representing bonds on the lattice, and typically only a few bonds are needed~\cite{PhysRevB.79.195125,Lichtenstein_2017,Profe_2022_tu2frg}. 

As a result we have $C(\mathbf{q}), D(\mathbf{q})$, and~$P(\mathbf{q})$ matrices in form-factor space. Superconductivity is signaled by the divergence of~$\abs{P}$, whereas a SDW is signaled by the divergence of~$\abs{C}$ (with~$\abs{D}$ tailing it with the ratio~$\abs{C} \approx 2\abs{D}$). The particular~$\mathbf{q}_*$ for which the channel component diverges indicates the ordering vector. By diagonalizing the matrix at~$\mathbf{q}_*$ in form-factor space, the gap symmetry can be read off the eigenvector associated to the diverging leading eigenvalue. Our calculations were performed with the publicly available \texttt{divERGe} code~\cite{Profe_2024_diverge,10.21468/SciPostPhysCodeb.26-r0.5}. Technical details and parameters used are given in the appendix.

\section{Analysis}
\subsection{Comparison of ambient- and high-pressure phases.} The band structure for the high-pressure $Fmmm$ phase and the ambient-pressure $Amam$ at $\delta \epsilon_*$ are shown in Fig.~\ref{Fig:1}(a,b). The Fermi surfaces for both phases are shown in Fig.~\ref{Fig:1}(c,d) at $\delta \epsilon^+$. While at a first glance the band structures appear quite distinct, we need to take into account that the unit cell of the $Amam$ phase is double that of the $Fmmm$ phase. 
Then, similarities become evident by folding the bands at the Brillouin zone center in the $Fmmm$ phase (the $M$ point) into the $\Gamma$ point of the $Amam$ model. Certain features can be easily identified between the models such as the two degenerate top-most bands around the $\Gamma$ point, the $\gamma$~band touching the Fermi level, and the $\beta$~band hovering between 1 and 2~eV. We note, however, that the bandwidths are slightly altered and appear a bit larger for the high-pressure model. This follows the expectation that increased pressure generically increases hopping amplitudes. 

\paragraph{Non-interacting susceptibilities} To elaborate, we compare the susceptibilities along a high-symmetry path for the pseudo-tetragonal system. Importantly, the susceptibility has structure factors which are dictated by the composition of the unit cell. Taking into account those, the overall features of the non-interacting susceptibilities in the ambient- and high-pressure phases are extremely similar, see for $\delta\epsilon^{+}$ (Fig.~\ref{Fig:1}(e) vs. Fig.~\ref{Fig:1}(f)), $\delta\epsilon^*$ (Fig.~\ref{Fig:1}(g) vs. Fig.~\ref{Fig:1}(h)) and $\delta\epsilon^-$ (Fig.~\ref{Fig:1}(i) vs. Fig.~\ref{Fig:1}(j)). The full information of the susceptibility for a freestanding bilayer is contained in the even and odd channels $\chi^{e/o} = \chi_\parallel \pm \chi_\perp$ \cite{Boetzel2024}. The even channel contains scatterings within the bonding and within the antibonding subspaces $\chi^e = \chi^{bb} + \chi^{aa}$. In contrast, scattering between bonding and antibonding subspaces contribute to the odd channel $\chi^o = \chi^{ba} + \chi^{ab}$. The most important observation is that the high-pressure and ambient-pressure curves are remarkably similar regarding the location of the peaks and their hierarchy. As the $\gamma$~pocket is moved downwards, the even strong low~$\mathbf{q}$ fluctuations associated to it are suppressed in agreement with~\cite{ryee2025superconductivity}, leaving the stage for the odd channel spin fluctuations. Remarkably, those are increased on the non-interacting level. As already pointed out in Ref.~\cite{ushio2025theoretical}, the largest contributions do not come from the Fermi surface but involve largely isotropic contributions from $\gamma-\delta$ scattering to $\chi^o_{zz}$. Only the peak related to $\gamma-\beta$ scattering which appears near the $X$ point is suppressed. On top of the largely isotropic $\gamma-\delta$ background, scatterings between $\alpha$ and $\beta$ bands give rise to an odd channel peak at $\mathbf{Q}_1$. At ambient pressure and with the $\gamma$ pocket being below the Fermi level, this peak is relatively close to the experimentally observed $(\pi/2,\pi/2)$ peak (pseudo-tetragonal notation), which would refer to ${\bf Q}_Y\approx(0,\pi)$ peak within the true orthorhombic $Amam$ unit cell (see Fig.~\ref{Fig:1}(d)). Recent inelastic neutron scattering unambiguously show that this peak is located at a fractional coordinate $L=2.5$ \cite{chen2026nature}, 
strongly suggesting that it must be associated with the odd channel \cite{Boetzel2024}. There is a second peak in the odd channel at/near $\mathbf{Q}_2 = (\pi,0)$ for the cases where the $\gamma$~band crosses or touches the Fermi level, which is subleading for the non-interacting susceptibility but was found to be important for the interacting susceptibilities \cite{yang2023possible,zhang2023structural,Boetzel2024}.

The last remark on the non-interacting susceptibility is that due to the larger bandwidth~$W$ at high pressure, the overall magnitude for the fluctuations is smaller compared than at ambient pressure.  
Since there are no significant differences between the high- and ambient-pressure phases in the non-interacting limit, we proceed studying the leading instabilities using the fRG.

\begin{figure}
	\includegraphics[width=1\linewidth]{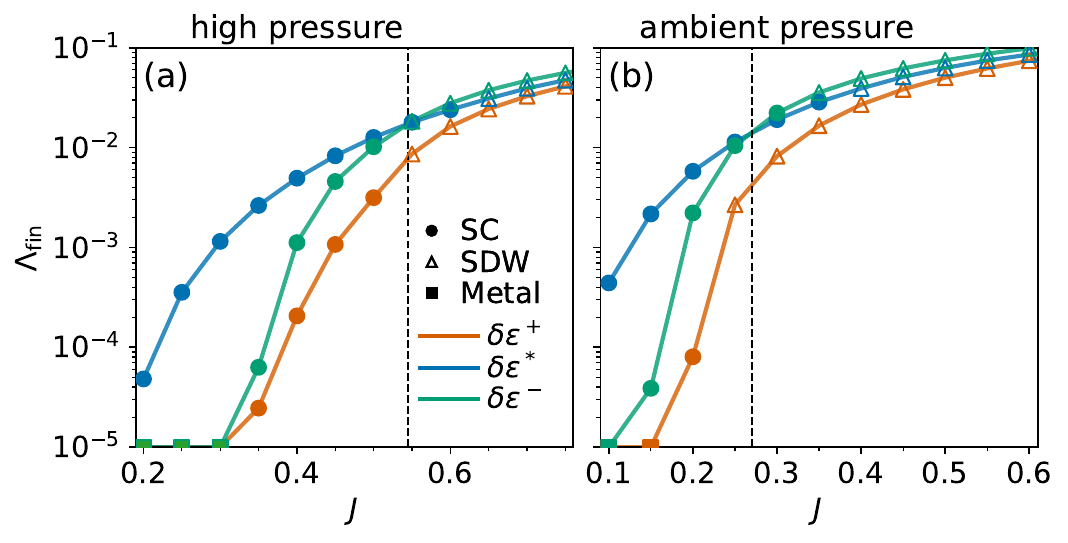}
	\caption{Leading instabilities within fRG for high- and ambient-pressure phases. Leading instabilities for the high~(a) and ambient~(b) pressure model as a function of Hund's coupling~$J$ for the $\gamma$~band being located 50 meV  above ($\delta\epsilon^+$), at ($\delta\epsilon^*$), and 50 meV below ($\delta\epsilon^-$) the Fermi level. Dashed black lines illustrate the transition from low $J$ to high $J$ regimes with the maximum $\Lambda_{\rm{fin}}$ changes from being at $\delta\epsilon^*$ to $\delta\epsilon^-$. }
\label{Fig:2}
\end{figure}

\paragraph{Leading instabilities from fRG}
We performed fRG calculations for the high-pressure and ambient-pressure models at $\delta \epsilon^\ast$ and $\delta\epsilon^\pm$. We set $U = 3$ eV and further varied $J$ within $0.1-0.8$ eV. The results are shown in Fig.~\ref{Fig:2}. We find a $s_{\pm}$ superconducting instabilities at lower $J$ and SDW instabilities in the odd channel with $\mathbf{Q}_{\rm{1}} = (\pi/2+\delta,\pi/2+\delta)$ at higher $J$. 

We can identify two different regimes, which are present in both, high- and ambient-pressure phases. 
(1)~At smaller~$J$, only superconducting instabilities are found and strongest for $\delta\epsilon^*$. This is in line with the estimated trend in Ref.~\cite{ryee2025superconductivity}, where the used $J/U$ is of comparable size. Closer analysis of the vertices shows that here two scattering vectors belonging to the odd channel are important $\mathbf{Q}_{\rm{1}}$ but also a second peak at/near $(\pi,0)$, see  Fig.~\ref{Fig:SM2} in the appendix. The latter contribution fades away when the $\gamma$ band is below the Fermi level. Fluctuations related to $\mathbf{Q}_1$ and $\mathbf{Q}_2$ work cooperatively to support the bonding-antibonding $s_\pm$ superconducting gap structure. 
(2)~In the higher~$J$ regime, the superconducting instability does not peak at $\delta\epsilon^*$. Instead, the instability is strengthened by increasing $\delta\epsilon$ further, so that the $\gamma$ band appears below the Fermi level. This behavior has also been seen using FLEX \cite{ushio2025theoretical}, where the used $J/U$ is again of comparable size. In this regime, the $\mathbf{Q}_{\rm{1}}$ contributions to the vertices become increasingly dominant. 

The transition to leading magnetic instability happens primarily in the high $J$ regime. The critical Hund's coupling $J_{\text{c}}$ at which the transition occurs is systematically higher for the high-pressure case. The transition does not show any obvious dependence on $\delta\epsilon$ and the position of the $\gamma$~band which is true for both phases. In particular, it is not just that there is a magnetic instability if the $\gamma$~band below the Fermi level and a superconducting instability once it is near/crossing the Fermi level. A more systematic variation of the crystal field splitting for the low and high $J$ regimes are compared exemplarily for the ambient pressure phase in Fig.~\ref{Fig:SM1} in the appendix . 

Concerning the character of the $s_{\pm}$ superconducting instabilities, it is always dominated by the interlayer $d_{z^2}$ term, so that the $s_{\pm}$ has to change sign between bonding and antibonding bands. The relative gap amplitudes are given in the appendix. The results for the high-pressure phase are consistent with earlier fRG studies~\cite{yang2023possible,Zhan2025fRG,le2025opposite}.
Decreasing $U$ was found to increase $J_\text{c}$ in both models. 
Interestingly, comparing $J_\text{c}$ in units of the respective bandwidth brings the high-pressure and ambient-pressure results closer together, which may hint that the differences in $J_\text{c}$ are at least partly driven by the change in bandwidth.

\begin{figure}
\includegraphics[width=1\linewidth]{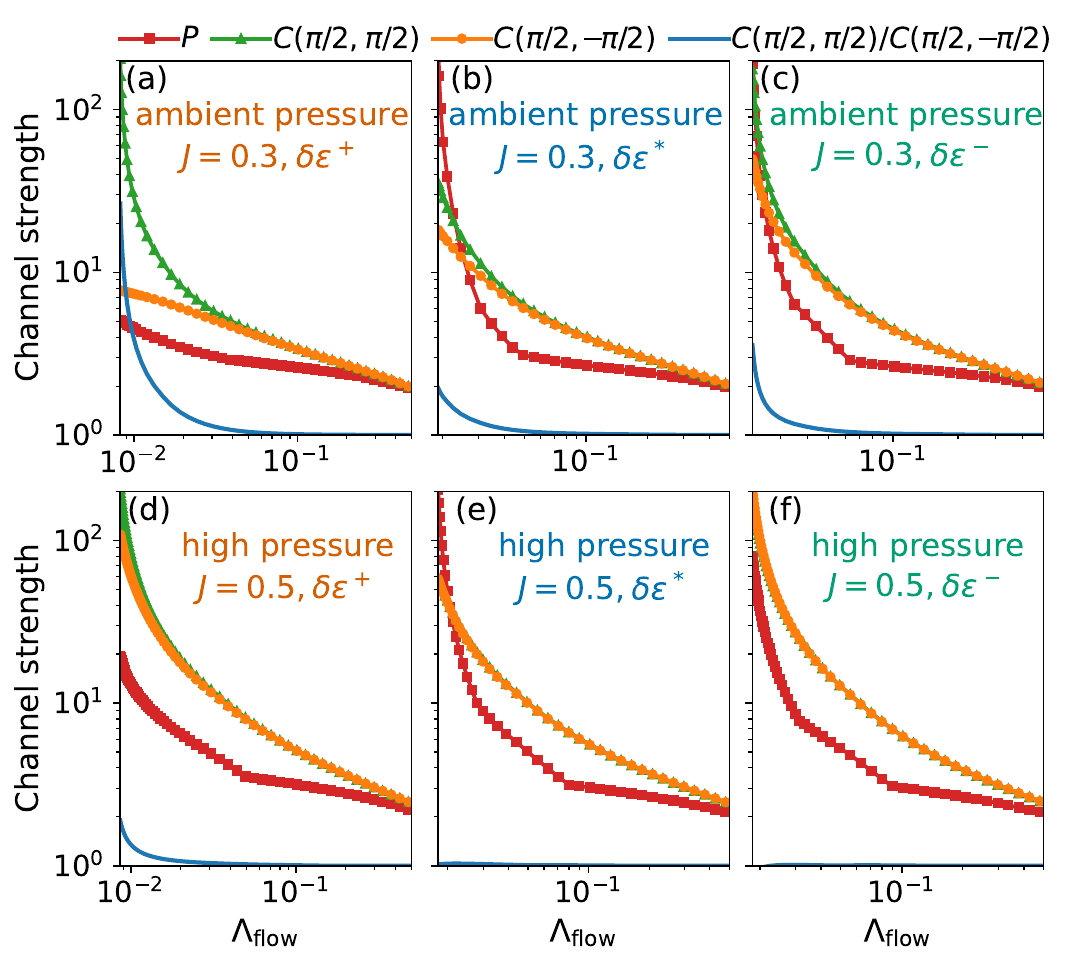}
	\caption{Flow of the maximal components of the particle-particle vertex $P$ and the crossed particle-hole vertex, i.e., $C(\mathbf{Q}_1 \approx (\pi/2,\pi/2))$ and $C(\mathbf{\tilde{Q}}_1\approx(\pi/2,-\pi/2))$ and their ratio. Flows are shown for $J=0.5$~eV for ambient pressure for the $\gamma$ band being located 50~meV  above ($\delta\epsilon^+$, (a)), at ($\delta\epsilon^*$ (b)), and 50 meV below ($\delta\epsilon^-$ (c)) the Fermi level. Panels (d), (e) and (f) display the corresponding flows for the high-pressure phase with $J = 0.3$ eV.}
\label{Fig:3}
\end{figure}

\subsection{Testing the Role of Orthorhombicity}
Despite the apparent similarities between the high-pressure and ambient pressure models outlined in the previous section, experiment show that superconductivity appears only at high pressures. It is difficult to explain this difference in the two phases in terms only of the bandwidths. Therefore, we now discuss the role of spin-fluctuations in the two phases in more detail. 

\paragraph{RG evolution of the vertex}
Spin-fluctuations, encoded in the particle-hole vertex $C$, in the two models drive superconductivity. We focus on a value $J$ near $J_\text{c}$ in the two regimes, namely $J = 0.3$ eV for the ambient-pressure phase and $J = 0.5$ eV for the high-pressure phase.  
To this end, in Fig.~\ref{Fig:3} we show the evolution of $C$ at the dominant~$\mathbf{Q}$ vectors and the particle-particle vertex~$P$ as a function of the RG scale $\Lambda$ for the two models for the three cases $\delta\epsilon^+$, $\delta\epsilon^*$ and $\delta\epsilon^-$. The two vectors $\mathbf{Q}_{1} \approx (\pi/2,\pi/2)$ and $\mathbf{\tilde{Q}}_{1} \approx (\pi/2,-\pi/2)$, which would be degenerate and related by related by a $C_4$ rotation in a tetragonal phase are the leading instabilities, while they would refer to ${\bf Q}_Y \approx (0,\pi)$ and ${\bf Q}_X \approx (\pi,0)$ in the orthorhombic $Amam$ unit cell. At ambient pressure, where orthorhombicity is significant, $C(\mathbf{Q}_{1})$ diverges first at the critical scale while $C(\mathbf{\tilde{Q}}_{1})$ lags behind. In contrast, the high-pressure phase is only slightly orthorombic for the considered $Fmmm$ space group there is little difference between $C(\mathbf{Q}_{1})$ and $C(\mathbf{\tilde{Q}}_{1})$. Interestingly, there is a clear correlation between the ratio of the two and the critical Hund's coupling $J_\text{c}$ at which the transition takes place.

\paragraph{Order-parameter degeneracy and fluctuation effects}
The conclusions we draw from Fig.~\ref{Fig:3} and our previous discussion are as follows. In the tetragonal high-pressure phase, there will generically be twice the number of magnetic order-parameter components, associated to the degenerate ordering vectors $\mathbf{Q}_1$ and ${\mathbf{\tilde Q}}_1$, which feed superconductivity when compared to the orthorhombic ambient-pressure situation.
In principle, the quasi two-dimensional nature of the layered material discourages ordering into a state that breaks a continuous symmetry due to strong fluctuations of the associated Goldstone modes, which destabilizes the order~\footnote{In perfect 2D, Goldstone fluctuations destabilize the ordered phase completely at finite temperature, as per the statement of the Hohenberg-Mermin-Wagner theorem.}. 
This tentatively plays against the formation of a SDW state, as the former typically features larger long-wavelength Goldstone fluctuations by virtue of SDW ordering vector having more order parameter components than the superconducting one. 
{While the fRG in the standard approximation employed here keeps track of the enhanced SDW fluctuations in the high-pressure phase feeding superconductivity, it does not}
properly take into account feedback with such Goldstone fluctuations and, in particular, it violates the Hohenberg-Mermin-Wagner theorem. 
{This suggests} that the role of orthorhombicity in making absent superconductivity is {even} underestimated in our fRG results. Finally, the degeneracy of the order parameter in the high-pressure case introduces frustration, which further discourages magnetic order. This instability is further reduced by the decreased commensurability of the dominant~$\mathbf{Q}_1$ and~${\mathbf{\tilde Q}}_1$ magnetic contributions in the high-pressure phase, which we observe in our calculations.

\begin{figure}
	\includegraphics[width=1\linewidth]{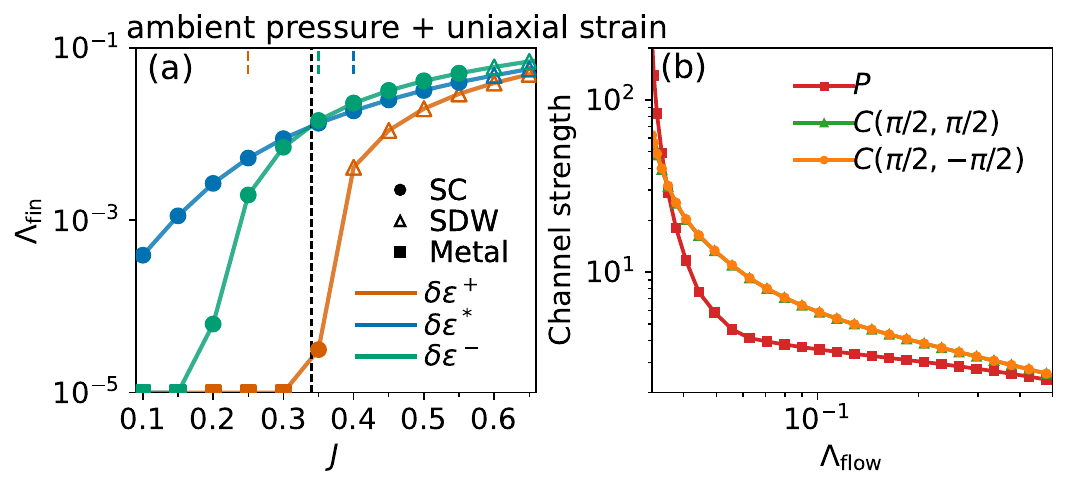}
	\caption{(a)~Leading fRG instabilities for ambient pressure with modeled uniaxial strain as a function of Hund's coupling $J$ for the $\gamma$~band being located 50~meV  above ($\delta\epsilon^+$), at ($\delta\epsilon^*$), and 50~meV below ($\delta\epsilon^-$) the Fermi level. Dashed black lines illustrate the transition from low~$J$ to high~$J$ regimes. Dashed colored lines have been added near the top of the panel to indicate the position of the transition to the SDW instability for ambient pressure in the absence of uniaxial strain.   (b)~Flow of~$P$ and the dominant contributions to~$C(\mathbf{{Q}}_{1} \approx (\pi/2,\pi/2))$ and $C(\mathbf{\tilde{Q}}_{1} \approx (\pi/2,-\pi/2))$ for $J = 0.5$~eV for $\delta\epsilon^*$ .}
\label{Fig:4}
\end{figure}



\paragraph{Making the ambient pressure phase tetragonal}
At this point, it is instructive to investigate the counter-hypothesis. We consider a model corresponding to ambient pressure but assume that application of uniaxial strain forces ($a=b$) within DFT (with relaxed atomic positions).  
The modeled strain brings the system closer to $C_4$~symmetry and also increases the overall bandwidth. The band structures for the different ambient pressure scenarios and their corresponding Fermi surfaces are shown in Fig.~\ref{Fig:SM3} in the appendix. For completeness, also the corresponding bare susceptibilities are shown, which are not strongly affected qualitatively. 

The impact of reducing the orthorhombicity on the leading fRG instabilities is visualized in Fig.~\ref{Fig:4}. 
We find that the pair $\mathbf{Q}_{1}$ and $\mathbf{\tilde{Q}}_{1}$ components of the magnetic vertex, which are related by $C_4$~rotation, grow closer together at low-energies, similar to what is observed to the high-pressure model. We note that the minor deviation from perfect $C_4$~symmetry is an artifact of the finite numerical thresholds chosen for the DFT calculations, which should mimic a more realistic experimental scenario where external stresses do not lead to perfect tetragonality. Therefore, we use a $C_4$~symmetrized version for Fig.~\ref{Fig:4}. A discernible difference from the ambient~pressure calculations of Fig.~\ref{Fig:2}(b) can be observed.

\section{Conclusion}
We investigated the competition between SDW and superconducting instabilities in the bulk bilayer nickelate $\mathrm{La_3Ni_2O_7}$ using functional renormalization group calculations. We demonstrated that the ambient-pressure ($Amam$) and high-pressure ($Fmmm$) phases are remarkably similar when comparing properties of their electronic band structures or their bare susceptibilities. This similarity suggests that the appearance of high-pressure superconductivity under pressure is difficult to explain purely by changes in the low-energy electronic structure between these two phases. 

To explain the experimentally observed differences, we propose the suppression of orthorhombicity as a key contributing factor to the appearance of the superconducting state in the high-pressure scenario. In particular, having a tetragonal or nearly-tetragonal structure introduces a near-degeneracy between symmetry-related SDW ordering vectors $\mathbf{Q}_{1}=(\pi/2,\pi/2)$ and $\mathbf{\tilde{Q}}_{1}=(\pi/2,-\pi/2)$. This effectively doubles the low-energy magnetic fluctuations that provide the pairing glue for the superconducting state. Simultaneously, the increase in the number of order-parameter components, as well as the resulting frustration, acts to further destabilize the formation of long-range magnetic SDW order, altogether shifting the balance in the high-pressure scenario toward the the formation of a superconducting state.

We supplemented the analysis by studying the effect of modeled uni-axial strain ($a=b$) in enhancing SDW fluctuations in a second ordering vector by reducing orthorhombicity. As an artifact of our numerical method, feedback from Goldstone fluctuations that should destabilize the SDW is absent. In the real system, however, we expect this to discourage the formation of an SDW state and the enhanced SDW fluctuations to support superconductivity. As an experimental pathway to test our conclusions, we propose that applying strain in the tetragonal high-pressure phase should kill or reduce superconductivity. Moreover, our analysis suggests that it may be possible to engineer high-$T_c$ superconductivity in the bilayer $\mathrm{La_3Ni_2O_7}$ at reduced or near-ambient pressures by leveraging uni-axial strain to reduce orthorhombicity.


\section{Acknowledgements} 
We thank Jonas B. Profe for helpful discussions and practical advice on the \texttt{divERGe} library, and Lauro B. Braz for helpful discussions.
The work is supported by the German Research Foundation Project No.572794210 (IME, FL, and SB). 
MMS acknowledges funding from the Deutsche Forschungsgemeinschaft (DFG, German Research Foundation) under Project No.~277146847 (SFB 1238, project C02) and Project No.~452976698 (Heisenberg program).

\bibliography{literature}

\clearpage
\onecolumngrid
\section{Appendix}
\subsection{Computational details and details on the superconducting instability}
\label{app:computational_details}
The DFT computations are performed within the local-density approximation using a mixed-basis pseudopotential framework~\cite{louie1979,elsaesser90,lechermann02,mbpp_code}. 
It utilizes norm-conserving pseudopotentials, as well as plane waves and atomic-like localized functions to represent the crystal wave functions. The plane-wave cutoff was chosen $E_{\rm cut}=16$\,Ryd and localized functions are introduced for La$(5d)$, Ni$(3d)$ and O$(2s,2p)$. A $k$-point mesh of size $9\times 9\times 9$ for $Fmmm$ and of size $5\times 5\times 5$ for $Amam$ is employed. The maximally-localized Wannier construction results in an effective low-energy Hamiltonian for the $\{d_{z^2},d_{x^2-y^2}\}$ orbitals from the Ni sites. 

We used a $100 \times 100$ momentum grid for the fRG flows at ambient pressure and $141 \times 141$ momentum grid for the high-pressure phase. The grids effectively result in the same momentum resolution because the folded Brillouin Zone is a factor of 2 smaller. To evaluate the bubbles an additional refinement factor of 4 is used for the momentum grids. For the presented data only intra unit cell bonds are used allowing better momentum resolution. However, we checked the role of higher order bonds to nearest and second nearest neighbors. The real space gaps for bonds to nearest neighbors are negligible. The gaps for the second nearest neighbors were found to be partially finite but without any qualitative impact on any of the presented results. The gap amplitudes are in agreement with a previous fRG study \cite{Zhan2025fRG}.

The relative gap amplitudes for the superconducting instabilities are listed exemplarily for the high-pressure phase in the in table \ref{tab:gap_amplitudes}. The values for ambient pressure scenarios are very similar. The interlayer $d_{z^2}$ term is always the dominant component with the dominance being largest for $\delta\epsilon^-$. This is expected because the odd susceptibility is most dominant for that case. 

\begin{table}[h]
\centering
\begin{tabular}{|c|c|c|c|}
\hline
J (eV) &  $\delta\epsilon=\delta\epsilon^+$ & $\delta\epsilon^*$ & $\delta\epsilon^-$ \\ 
\hline
0.2 &   - & $\begin{pmatrix}
-0.448 &  0.100 \\
0.537 &  -0.014
\end{pmatrix}$ & -  \\ \hline
0.4 &   $\begin{pmatrix}
-0.449 & 0.120 \\
0.524 & 0.095
\end{pmatrix}$ & $\begin{pmatrix}
-0.394 &  0.094 \\
0.577 &  0.055
\end{pmatrix}$ & $\begin{pmatrix}
-0.109  & 0.115 \\
 0.676 &  0.132
\end{pmatrix}$  \\ \hline
0.5 &   $\begin{pmatrix}
-0.437 & 0.110 \\
0.534 & 0.112
\end{pmatrix}$ & $\begin{pmatrix}
-0.380 &  0.088 \\
0.585 &  0.074
\end{pmatrix}$ & $\begin{pmatrix}
-0.193  & 0.010 \\
 0.659 &  0.135
\end{pmatrix}$  \\ \hline
0.6 & -
& $\begin{pmatrix}
-0.368 & 0.083 \\
0.519 & 0.088
\end{pmatrix}$ & -
\\ \hline
\end{tabular}
\caption{Relative real-space gap amplitudes for the superconducting instability for different $J$ in the format $\begin{pmatrix}
\Delta_\parallel^z  & \Delta_\parallel^x \\
 \Delta_\perp^z &  \Delta_\perp^x
\end{pmatrix}$ shown for the high-pressure phase, where $\Delta_\parallel$ denotes the on-site component. The square sum of all gap components is normalized to one.}
\label{tab:gap_amplitudes}
\end{table}

\subsection{Details on the role of Hund's coupling}
\begin{figure}[h]
	\includegraphics[width=1\linewidth]{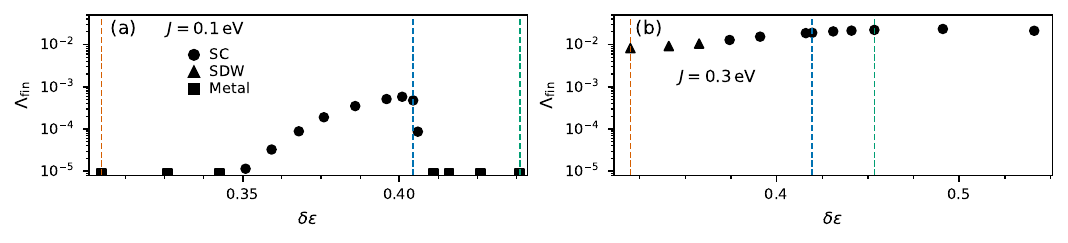}
	\caption{Leading instabilities within fRG for ambient pressure phase. Scan of $\delta \epsilon$ for the ambient-pressure phase at low $J = 0.1$ eV (c) and high $J = 0.3$ eV (d) with dashed lines denoting $\delta\epsilon^+$, $\delta\epsilon^*$ and $\delta\epsilon^-$ values.}
\label{Fig:SM1}
\end{figure}

In Fig.~\ref{Fig:SM1} the critical flow parameters for the different instabilities are shown as a function of $\delta\epsilon$ for the ambient pressure phase inside the low $J$ and high $J$ regime. In the low $J$ regime, the maximum is located in vicinity of $\delta\epsilon^*$. As soon as the $\gamma$ pocket is shifted below the Fermi level, the instability is quickly suppressed whereas there is a slower suppression when $\delta\epsilon$ is enhanced. In contrast, there is relatively little dependence on $\delta\epsilon$ at higher $J$. Especially, there is no maximum near $\delta\epsilon^*$.

\begin{figure}
	\includegraphics[width=1\linewidth]{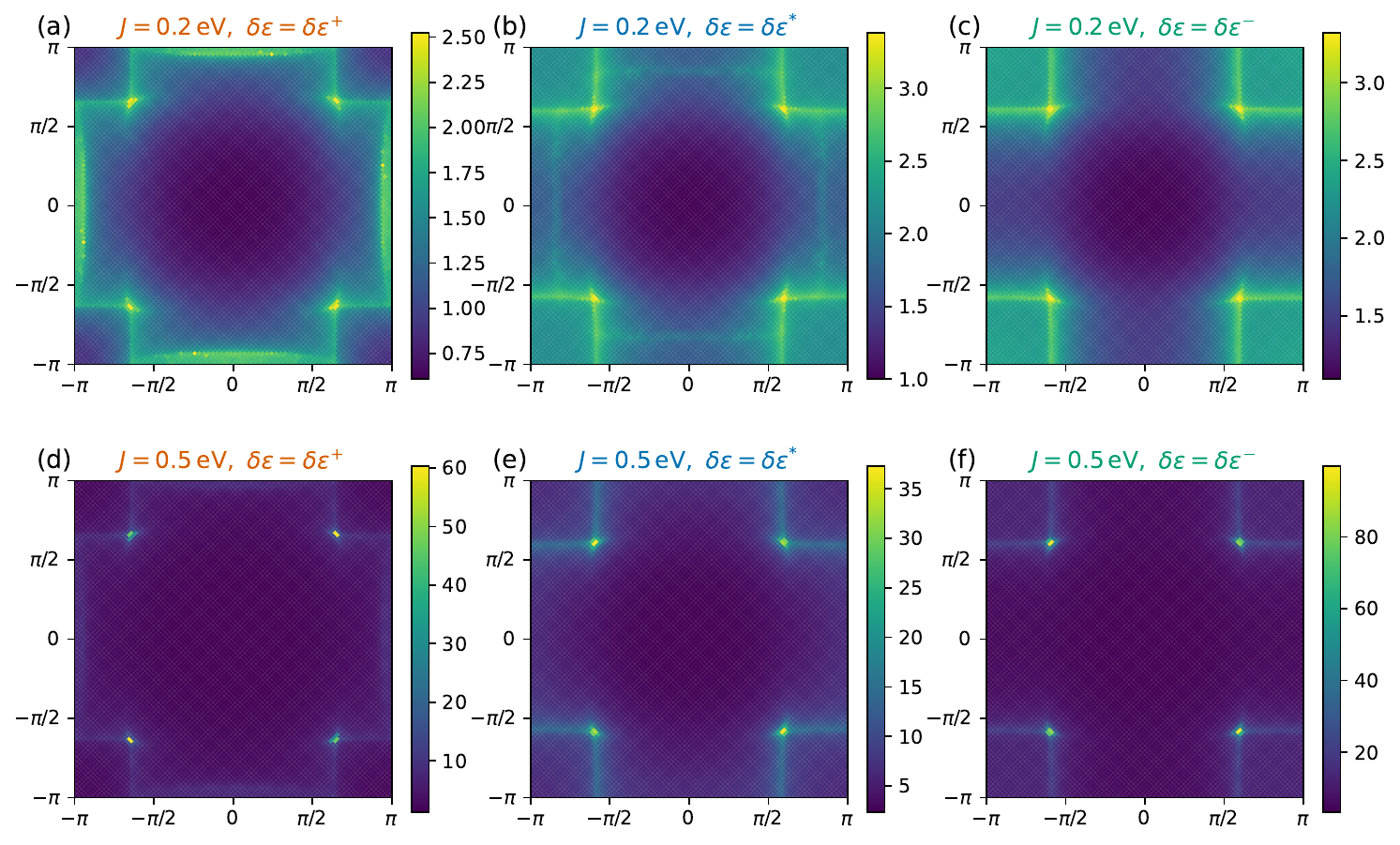}
    \caption{Magnetic susceptibilities in the high-pressure phase at the critical flow parameters in the high and low $J$ regimes, shown for different locations of the $\gamma$ band. Panels (a)-(c) represent the low $J$ regime with $J=0.2$ eV for $\delta\epsilon=\delta\epsilon^+$~(a), $\delta\epsilon^*$~(b) and $\delta\epsilon^-$~(c). Panels (d)-(f) show the corresponding results representing the high $J$ regime with $J=0.5$ eV.}
\label{Fig:SM2}
\end{figure}

This behavior can be understood by considering the magnetic susceptibilities, which are shown in Fig.~\ref{Fig:SM2}. In the low $J$ regime, we are further away from $J_\text{c}$. For $\delta\epsilon^*$ and $\delta\epsilon^+$ there are significant contributions not only near $\mathbf{Q}$ and $\mathbf{\tilde{Q}}$ but also from near $(\pi,0)$ and $(0,\pi)$. These contributions seemingly cooperatively cause the superconducting instability. As the latter contributions disappear once the $\gamma$ pocket is below the Fermi level, the superconducting instability also disappears. At higher $J$, contributions from $\mathbf{Q}$ and $\mathbf{\tilde{Q}}$ but also from near $(\pi,0)$ and $(0,\pi)$. These contri are large and very dominant so that the role of other contributions becomes irrelevant.

\clearpage

\subsection{Band structure and bare susceptibilities for ambient pressure with uniaxial strain}
\begin{figure}[h]	                 \includegraphics[width=1\linewidth]{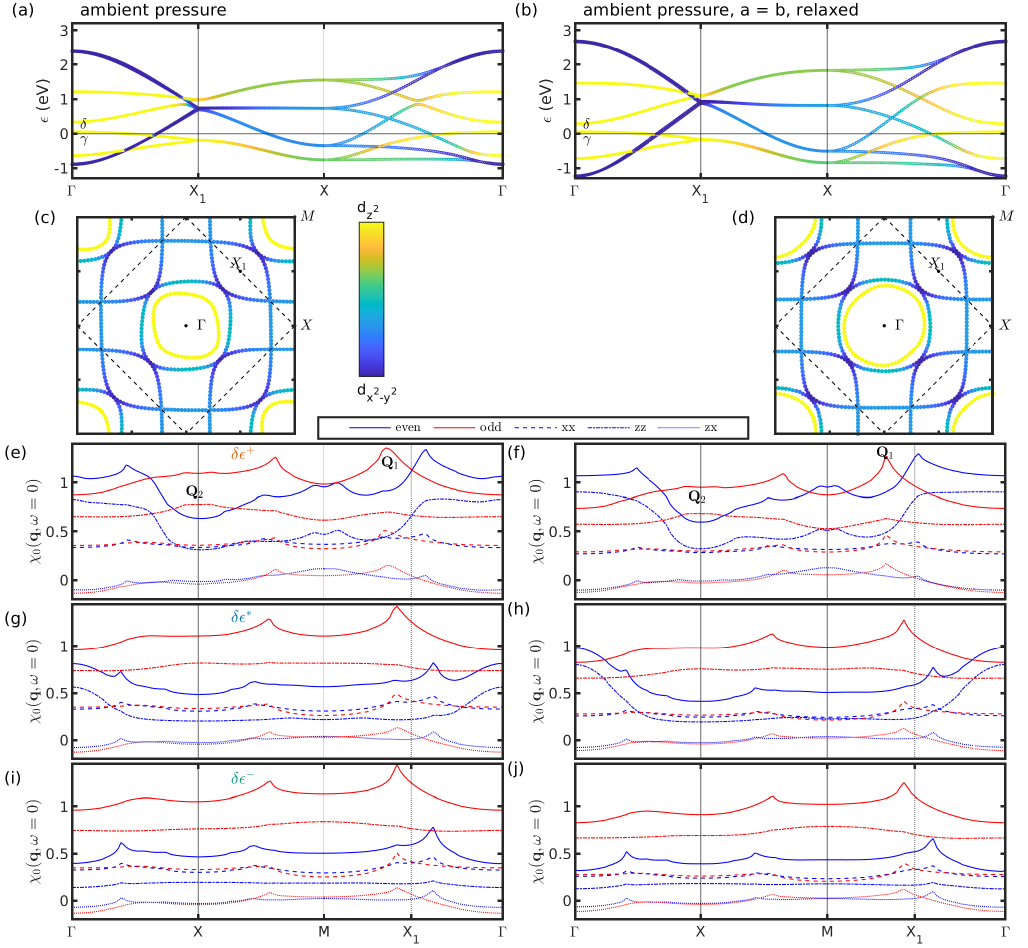}
    \caption{Band structure, Fermi surface and non-interacting susceptibility of La$_3$Ni$_2$O$_7$ at ambient pressure with modeled strain. Band structure for the ambient pressure model along $Amam$ high symmetry path for $\delta\epsilon = \delta\epsilon^*$ without strain (a) and with strain (b). (c-d) The Fermi surfaces corresponding to the two cases for $\delta\epsilon = \delta\epsilon^+$. (e)-(j) Total and orbital resolved contributions to the non-interacting susceptibilities for the top of the $\gamma$ band for $\delta\epsilon^+$, $\delta\epsilon^*$ and $\delta\epsilon^-$ for the three cases. The ambient pressure data for the case without strain is the same as in Fig.~\ref{Fig:1} and repeated here for an easier comparison.}
\label{Fig:SM3}
\end{figure}

The band structure for the ambient pressure model (Fig.~\ref{Fig:SM3}(b)) with strain is compared to the ambient pressure model (Fig.~\ref{Fig:SM3}(a)). The strain enlarges the bandwidth from $3.35$ eV to $3.92$ eV for $\delta\epsilon=\delta\epsilon^*$ (the high-pressure bandwidth is $4.11$ eV). The Fermi surface for uniaxial strain model (Fig.~\ref{Fig:SM3}(d)) is almost $C_4$ symmetric.

\end{document}